\documentclass[pra,twocolumn,groupedaddress,showpacs,floatfix]{revtex4-2}

\usepackage{graphics}
\usepackage{graphicx}
\usepackage{bm}
\usepackage{amsmath,accents}
\usepackage{amsfonts}
\usepackage{amssymb}
\usepackage{latexsym}
\usepackage{color}
\usepackage{bbold}
\usepackage{braket}
\usepackage{float}
\usepackage{hyperref}
\usepackage{extarrows}
\usepackage{dsfont}
\usepackage{subfigure}
\usepackage[rflt]{floatflt}

\begin{document}

\title{Comparing Hermitian and Non-Hermitian Quantum Electrodynamics}
\author{Jake Southall, Daniel Hodgson, Robert Purdy and Almut Beige}
\affiliation{The School of Physics and Astronomy, University of Leeds, Leeds LS2 9JT, United Kingdom}

\date{\today}

\begin{abstract}
In recent years, non-Hermitian quantum physics has gained a lot in popularity in the quantum optics and condensed matter communities in order to model quantum systems with varying symmetries. In this paper, we identify a non-standard inner product that implies bosonic commutator relations for local electric and magnetic field observables and leads to a natural local biorthogonal description of the quantised electromagnetic field. When comparing this description with an alternative local Hermitian description, in which the states of local photonic particles -- i.e.~of so-called bosons localised in position (blips) -- are orthogonal under the conventional Hermitian inner product, we find that there is an equivalence between the two approaches. Careful consideration needs to be given to the physical interpretation of the different descriptions. Whether a Hermitian or a non-Hermitian approach is more suitable depends on the circumstances that we want to model.
\end{abstract}

\maketitle 

\section{Introduction}

In classical electrodynamics, the fundamental equations of motion are Maxwell's equations: a set of  highly-symmetric differential equations that describe the relationship between the local electric and magnetic fields defined at each point in {\em space and time}. In contrast to this, in quantum electrodynamics, we routinely decompose the local electromagnetic (EM) field into objects called photons which have a well-defined {\em momentum} that does not change in time. Due to the Heisenberg uncertainty principle, monochromatic photons have a maximally undefined location. A key reason for expressing the EM field in this way is that monochromatic photons are energy eigenstates and thus valuable for modelling scenarios regarding energy conservation. For example, when an atom absorbs a photon at its resonance frequency, it will be excited to a higher energy state separated from the initial state by a well-defined amount of energy. 

As recently emphasized, for example in Refs.~\cite{Daniel}, the symmetries of monochromatic photons are too restrictive making them insufficient to describe all possible wave packets allowed by classical electrodynamics. For example, the solutions of Maxwell's equations include highly-localised wave packets which propagate at the speed of light without dispersion. But when we superpose the monochromatic photon states that are allowed by standard quantum electrodynamics \cite{EJP} to form a highly localised wave packet, dispersion cannot be avoided. Non-local descriptions of light \cite{Scheel, Buhmann, Philbin,Gruner} therefore make it challenging, for example, to model the dynamics of electric and magnetic field vectors in the presence of optical elements, which are highly-localised objects. In the literature, many authors have therefore discussed possible local quantisations of light and, in particular, considered how to construct local bosonic excitations of the EM field (see e.g.~Refs.~\cite{Raymer, Hawton_2017, BB, Sipe, Cook, Jake, Daniel,Casimir} and references therein).  These local excitations, like the monochromatic photon states we are used to, can be created and annihilated by a set of creation and annihilation operators that commute at all non-zero displacements.

Nevertheless, it is often convenient to work with states of well-defined energy, since they have a particularly useful property. To see this, let us examine Ehrenfest's theorem, which describes how expectation values evolve in time. When an operator $A$ has no implicit time dependence, Ehrenfest's theorem states that the time derivative of the expectation value of $A$ with respect to a time-dependent state $\ket{\psi(t)}$ is simply given by
\begin{eqnarray}\label{H equation}
\frac{\rm d}{{\rm d}t}\bra{\psi(t)} A \ket{\psi(t)} &=&-\frac{i}{\hbar}\bra{\psi(t)}[A,H]\ket{\psi(t)} \, , 
\end{eqnarray}
where $H$ is the Hermitian Hamiltonian of the system. This equation shows that the dynamics of expectation values is intimately linked to the commutator between the Hamiltonian and other operators. When both the operator $A$ and the Hamiltonian $H$ are expressed in terms of bosonic operators, calculations of time derivatives of expectation values become very straightforward. Clearly, utilising operators with bosonic commutator relations is in general highly advantageous. For this reason, it is typically the monochromatic photon operators that are used when constructing observables. 

Unfortunately, as mentioned above, monochromatic photons are non-local which makes modelling interactions with localised objects challenging. For example, in theoretical descriptions based on the standard inner product of quantum physics, the commutator of the local field observables, ${\bf E}$ and ${\bf B}$, with the system Hamiltonian, $H$, is not simple. In this paper, we therefore consider an alternative representation of the EM field in terms of local Fock-space excitations that belong to a biorthogonal system \cite{Brody, Brody_2016}: a type of system that arises in non-Hermitian physics and which uses a non-standard inner product. The main advantage of this description is that its local electric and magnetic field vectors are pairwise orthogonal and commute simply with the field Hamiltonian, $H$; this is valuable in situations where the focus is on the dynamics of these vectors. In contrast, in our previous papers \cite{Jake,Daniel,Casimir}, we quantised the EM field in terms of pairwise orthogonal local energy quanta, so-called bosons localised in position (blips), which are well-suited for modelling the dynamics of localised particles.

As was explored in Ref.~\cite{Daniel}, the exact nature of these blips is determined by the symmetry group of the space in which they are contained. In particular, in free space, it is the translation symmetry of the Poincar\'e group that leads to the simple form of the blips' equation of motion. If the symmetry group of the considered space is reduced, this has the effect of altering the equation of motion in some regions of the space \cite{Casimir}. For example, when considering an optical cavity, the mirrors forming the walls of the cavity break the full translation invariance, leading to an altered motion for the blips in the location of the mirror. Similarly, in order to model the interactions of an atom with the EM field, our space becomes a \textit{pointed space}, with the location of the atom promoted to a distinguished position. In this case, we again clearly lose translation invariance but -- depending on the sophistication of the model -- potentially retain the rotational symmetries of the Poincar\'e group. As with the cavity, it is the particular set of reduced symmetries that determines the motion of field excitations, and hence the behaviour of the system.

As in Refs.~\cite{Hawton_2017,Cook, Jake, Daniel,Casimir}, in this paper, we need to make an adjustment to the standard theory of the quantised EM field. Specifically, we need to extend the Hilbert space of monochromatic modes to include all modes with a frequency in the range $(-\infty, \infty)$. This means we do not restrict ourselves to positive-frequency photons but allow photon frequencies to be positive and negative. This adjustment ensures that localised excitations in one dimension that have a clear direction of propagation do not disperse. This is in good agreement with classical electrodynamics, where wave packets with a well-defined direction of propagation also travel at the speed of light without changing their shape \cite{Jake, Daniel}. Moreover, the extension of the frequency range of the photons ensures that our local description of light does not violate any no-go theorems concerning the localisability of the photon, which have been put forward by several authors \cite{Pryce, Wigner, Wightman, Hegerfeldt, Halvorson, Malament, Knight}.

In the following, we compare the predictions of the local Hermitian description of the quantised EM field, which we introduced in Refs.~\cite{Jake,Daniel,Casimir}, with the non-Hermitian local description that we introduce here, and highlight the connection between the two approaches. Systems possessing curious non-Hermitian structures have gained much interest in recent years \cite{Bender,Ali}, particularly in quantum optics \cite{ElGanainy, ElGanainy2, Hawton_2017, Raymer}. For example, Hawton's and Debierre's approach \cite{Hawton_2017} uses biorthogonal quantum mechanics. However, their approach also uses a time-dependent inner product in an interaction picture. The inner product we use in this paper is not time-dependent and can be used in any picture. 

There are five sections in this paper. In Section 2, we shall cover some requisite background material on the quantised EM field and biorthogonal quantum mechanics. In Section 3, we shall model the EM field using biorthogonal quantum mechanics, and in Section 4 we shall describe how this biorthogonal approach connects to our earlier Hermitian approach. Finally, we present our conclusions in Section 5. Some mathematical details have been placed in Appendices A-C to simplify the reading of the manuscript.

\section{Theoretical background}

In this section, we introduce the theoretical background and concepts used throughout the rest of the paper. Before examining biorthogonal quantum mechanics and pseudo-Hermitian physics, we first review the standard description of the quantised EM field, which can be found in many quantum optics textbooks (see e.g.~Ref.~\cite{EJP} and references therein).

\subsection{The quantised EM field} \label{sec21}

The classical theory of electromagnetism describes the evolution of two fundamental quantities: the electric field and the magnetic field, $\textbf{E}(\bm{x},t)$ and $\textbf{B}(\bm{x},t)$. The dynamics of these fields is governed by Maxwell's equations, which take the form
\begin{eqnarray}\label{Maxwell}
&& \nabla \cdot \textbf{E}(\bm{x},t) = 0\,, ~~ \nabla \cdot \textbf{B}(\bm{x},t) = 0\,, \nonumber\\
&& \nabla \times \textbf{E}(\bm{x},t) = -\frac{\partial}{\partial t}\textbf{B}(\bm{x},t)\,, \nonumber\\
&& \nabla \times \textbf{B}(\bm{x},t) =  \frac{1}{c^2}\frac{\partial}{\partial t}\textbf{E}(\bm{x},t)
\end{eqnarray}
in the absence of any charges or source currents. The classical Hamiltonian $H_{eng}$ of the free-space EM field is 
\begin{equation}\label{classical energy}
H_{eng}=\int_{\rm \bf V} {\rm d}{\rm \bf V} \left(\epsilon_0{\rm \bf E}({\bf x}, t)^2+\frac{1}{\mu_0}{\rm \bf B}({\rm \bf x}, t)^2\right)\,,
\end{equation}
where ${\rm \bf V}$ denotes the volume enclosing the EM field, while $c$ is the speed of light and $\epsilon_0$ and $\mu_0$ are the permittivity and the permeability of free space.

One way of quantising the EM field created by light travelling along the $x$-axis is to assume that its basic building blocks are monochromatic photons \cite{EJP}. Adopting the notation which we employ also later in this paper, we can model these monochromatic photons by using the bosonic Fock ladder operators, $a_{s\lambda}(k)$ and $a^{\dagger}_{s\lambda}(k)$, where $s=\pm1$ denotes the direction of propagation along the $x$-axis, $\lambda =1,2$ denotes their polarisation and $\omega = ck$ denotes their frequency, with $k \in (0,\infty)$. These operators satisfy the bosonic commutation relations
\begin{gather}
\label{com1}
\left[a_{s\lambda}(k), a_{s'\lambda'}(k')\right]= \left[a^{\dagger}_{s\lambda}(k), a^{\dagger}_{s'\lambda'}(k') \right] = 0\,,\nonumber\\
\left[a_{s\lambda}(k), a^{\dagger}_{s'\lambda'}(k') \right] = \delta_{ss'}\delta_{\lambda\lambda'}\delta(k-k')\,.
\end{gather}
Using these operators we can create a single-photon state by acting a creation operator on the vacuum state $|0\rangle$,
\begin{equation}
\ket{1_{s\lambda}(k)}=a^{\dagger}_{s\lambda}(k)\ket{0}\,,
\end{equation}
where the vacuum state satisfies $\braket{0|0}=1$ and $a_{s\lambda}(k)\ket{0}=0$ for all $s$, $k$ and $\lambda$. Clearly, monochromatic single-photon states are pairwise orthogonal, since
\begin{equation} \label{6}
\braket{1_{s\lambda}(k)|1_{s'\lambda'}(k'}= \left[a_{s\lambda}(k), a^{\dagger}_{s'\lambda'}(k') \right] = \delta_{ss'}\delta_{\lambda\lambda'}\delta(k-k')\,.
\end{equation}
We can construct further states of the quantised EM field by acting on $|0\rangle$ with multiple creation operators. 

For light propagating only along the $x$-axis, the field observables for the quantised EM field at a point $x$ can be written as \cite{EJP}
\begin{eqnarray} \label{localfields}
 {\bf E}(x)&=&\sum_{s,\lambda} \sqrt{\frac{\hbar c}{4\pi\varepsilon A}} \, \int_0^\infty {\rm d} k \, \sqrt{k} \, {\rm e}^{i skx} \, a_{s\lambda}(k) \,  {\bf e}_\lambda+ H.c. \, , \nonumber \\
 {\bf B}(x)&=&\sum_{s,\lambda} {s \over c} \, \sqrt{\frac{\hbar c}{4\pi\varepsilon A}} \, \int_0^\infty {\rm d} k \, \sqrt{k} \, {\rm e}^{{\rm i} skx} \, a_{s\lambda}(k) \,  {\bf e}_x\times {\bf e}_\lambda  \nonumber \\ 
 &&+ H.c. \, , 
\end{eqnarray}
where $H.c.$ denotes the Hermitian conjugate and where the $\textbf{e}_\lambda$ are unit vectors oriented in the $y$ and $z$ directions. The Hamiltonian that generates the free-space dynamics of light is given by
\begin{equation}\label{freeH}
H_{eng}=\sum_{s,\lambda}\int_0^{\infty}{\rm d}k\, \hbar c k\,a^{\dagger}_{s\lambda}(k)a_{s\lambda}(k)
\end{equation}
up to a constant term -- the zero point energy -- which does not contribute to the dynamics. Under this Hamiltonian, field expectation values evolve as predicted by Maxwell's equations. However, as mentioned above, the $a_{s\lambda}(k)$ operators correspond to non-local photons and therefore do not always provide an intuitive description.

\subsection{Biorthogonal quantum mechanics}

Let us now review some relevant properties of biorthogonal quantum mechanics. A defining feature of a Hilbert space is its inner product. Since this paper uses different inner products, we denote the inner product between two states on different Hilbert spaces in the following by $\braket{\ket{\psi}, \ket{\phi}}^{ss}$, with the superscript labelling the particular inner product. For simplicity, the conventional inner product will be denoted with no superscript. The expectation value of an operator $A$ with respect to a state $\ket{\psi}$ under the conventional inner product is
\begin{equation}
\braket{A\ket{\psi}, \ket{\psi}}=\bra{\psi}A\ket{\psi}\,.
\end{equation}
In the following, we use the term ``non-Hermitian operators'' to refer to operators that are non-Hermitian with respect to the conventional inner product. However, note that this does not necessarily mean they are non-Hermitian with respect to a different inner product. We reserve the dagger notation $\dagger$ to denote the Hermitian adjoint of an operator with respect to the conventional inner product.

Suppose a set of $N$ linearly independent states $\{\ket{\alpha_n}\}$ for $n \in \{1, ..., N\}$ spans an $N$-dimensional Hilbert space, but is not necessarily orthonormal with respect to the conventional inner product. Then it is possible to obtain a set of $N$ states $\{\ket{\beta_n}\}$ such that $\braket{\beta_i|\alpha_j}=\delta_{ij}$. To see this, for a given $\ket{\alpha_m}$ one can select a state $\ket{\beta_m}$ from the one-dimensional subspace of the Hilbert space orthogonal to the span of $\{\ket{\alpha_n}\}_{n\ne m}$ such that $\braket{\beta_m|\alpha_m}= 1$. If we continue this process for all $\ket{\alpha_n}$, we can construct the set $\{\ket{\beta_n}\}$ \cite{Brody_2016}. The set $\{\ket{\beta_n}\}_{n=1}^{N}$ is called the biorthonormal basis associated with $\{\ket{\alpha_n}\}_{n=1}^{N}$. Given a state 
\begin{equation}\label{phi state}
    \ket{\psi}=\sum_n a_n\ket{\alpha_n},
\end{equation}
we define an associated state
\begin{equation}
    \ket{\widetilde{\psi}}=\sum_n a_n\ket{\beta_n} \, .
\end{equation}
The biorthonormal quantum mechanical (BQM) inner product can then be defined on this Hilbert space as
\begin{equation}
    \braket{\ket{\psi_1},\ket{\psi_2}}^{BQM}=\braket{\widetilde{\psi}{}_2|\psi_1}\,.
\end{equation}
Under this inner product, the set $\{\ket{\alpha_n}\}_{n=1}^{N}$ forms an orthonormal basis.  Furthermore, operators of the form  
\begin{equation}\label{hermitian operator}
	A=\sum_{n,m}a_{nm}\ket{\alpha_n}\bra{\beta_m}
\end{equation}
for real $a_{nm}$ are Hermitian with respect to the BQM inner product and are therefore known as biorthogonally Hermitian operators.

Collectively the set of states $\{\ket{\alpha_n},\ket{\beta_n}\}_{n=1}^{N}$ constitutes a biorthogonal system.  In the literature on biorthogonal quantum mechanics,  it is shown that if a state $\ket{\psi}$ belongs to a Hilbert space $\mathcal{H}$, its associated state $\ket{\widetilde{\psi}}$ is said to belong to the dual Hilbert space $\mathcal{H}^*$.  To avoid confusion with other definitions of a dual space, we refer to $\mathcal{H}^*$  in this paper as the bio-conjugate Hilbert space.  Hence, for every state $|\psi\rangle$ in Hilbert space, there exists a state $|\widetilde{\psi}\rangle$ in the bio-conjugate space. In the case that $\{\ket{\alpha_n}\}_{n=1}^{N}$ already forms an orthonormal basis, selecting $\ket{\beta_n}=\ket{\alpha_n}$ reduces the BQM inner product to the standard inner product, and the Hilbert space and bio-conjugate space become the same.  

\subsection{Pseudo-Hermitian Physics}

Biorthogonal quantum physics is closely related to pseudo-Hermitian physics. To see that this is so, suppose that we have a Hamiltonian $H$ acting on an $N$-dimensional Hilbert space $\mathcal{H}$.  This Hamiltonian is said to be pseudo-Hermitian if it satisfies the relation 
\begin{equation}
	\label{PH1}
    H^\dagger=\eta H\eta^{-1} 
\end{equation}
for some invertible operator $\eta$ satisfying $\eta=\eta^\dagger$ also acting on $\mathcal{H}$.  If one defines an inner product $\langle \cdot |\cdot \rangle^\eta$ such that
\begin{equation}\label{eta inner product}
	\braket{\ket{\psi_1},\ket{\psi_2}}^{\eta}=\bra{\psi_2}\eta\ket{\psi_1}\,,
\end{equation}
then by using Eq.~(\ref{PH1}) one may show that
\begin{equation}
    \braket{H\ket{\psi_1},\ket{\psi_2}}^{\eta}=\braket{\ket{\psi_1},H\ket{\psi_2}}^{\eta}\,.
\end{equation}
Because this is the definition of Hermiticity,  it follows that a pseudo-Hermitian Hamiltonian $H$ satisfying Eq.~(\ref{PH1}) is Hermitian in the proper sense with respect to the $\eta$-inner product given in Eq.~(\ref{eta inner product}).  

It is well known that the eigenstates of a Hermitian operator are orthogonal and that their eigenvalues are real. Because $H$ is Hermitian with respect to the $\eta$-inner product, the eigenstates of $H$ can also be shown to be orthogonal to one another under the same inner product. Assuming $H$ to be non-degenerate, let us denote the set of $N$ normalised and orthogonal eigenstates of $H$ by $\{|\alpha_n\rangle\}^{N}_{n = 1}$.  
If we define the states 
\begin{equation}
\label{beta1}
\ket{\beta_n}=\eta\ket{\alpha_n}
\end{equation}
then we can see by taking the $\eta$-inner product between $\alpha_n$ states, and by using the orthogonality of such states that
\begin{equation}
\langle \alpha_n|\eta|\alpha_m\rangle = \langle \beta_n|\alpha_m\rangle = \delta_{nm}\,.
\end{equation} Hence $\{\ket{\alpha_n},\ket{\beta_n}\}_{n=1}^{N}$ describes a biorthonormal system just as was discussed previously. Moreover,
\begin{equation}
    I_d=\sum_n^N\ket{\alpha_n}\bra{\beta_n}
\end{equation}
represents the identity operator for such a system.

Using the definition in Eq.~(\ref{beta1}) one can further show that, whereas the $|\alpha_n\rangle$ states are orthonormal with respect to the $\eta$-inner product, the $|\beta_n\rangle$ states are orthonormal with respect to the $\eta^{-1}$ inner product,
\begin{equation}\label{eta^{-1} inner product}
	\braket{\ket{\psi_1},\ket{\psi_2}}^{\eta^{-1}}=\bra{\psi_2}\eta^{-1}\ket{\psi_1}\,.
\end{equation}
So, if $\{\ket{\alpha_n},\ket{\beta_n}\}_{n=1}^{N}$ is a biorthonormal system equipped with the $\eta$-inner product, then $\{\ket{\beta_n},\ket{\alpha_n}\}_{n=1}^{N}$ is a biorthonormal system equipped with the $\eta^{-1}$-inner product. Note that this means these sets of states, while belonging to the same vector space, do not belong to the same \textit{Hilbert space}. Since a defining feature of a Hilbert space is its inner product, if we define the inner product between the $\ket{\alpha_n}$ states to be the $\eta$ inner product, and the inner product between the $\ket{\beta_n}$ states to be the $\eta^{-1}$ inner product, then the two Hilbert spaces are distinct. Unlike the $|\alpha_n\rangle$ states, the $|\beta_n\rangle$ states are not in general also orthonormal with respect to the $\eta$-inner product.  To see this simply compute
\begin{equation}\label{gamma not orthonormal}
	\braket{\ket{\beta_n},\ket{\beta_m}}^{\eta}=\bra{\alpha_m}\eta^3\ket{\alpha_n} \not\equiv\delta_{nm}\,.
\end{equation}
Note that, if $\eta$ is the identity operator on $\mathcal{H}$, Eq.~\eqref{eta inner product} reduces to the conventional inner product of quantum mechanics.

Eq.~\eqref{gamma not orthonormal} is not a problem if the only states of concern are those normalised with respect to the $\eta$-inner product, and if the only observables of concern are of the form given in Eq.~\eqref{hermitian operator}. In this case, one can select the $\eta$-inner product for the Hilbert space and will find that pseudo-Hermitian quantum mechanics is indistinguishable from conventional Hermitian quantum mechanics \cite{Brody_2016}. Likewise, if all states of concern are those normalised with respect to the $\eta^{-1}$-inner product and all observables are Hermitian with respect to this inner product, one can simply select the $\eta^{-1}$-inner product for their Hilbert space. 

\subsection{Time-evolution}

\begin{figure}[t]
\centering
\includegraphics[width=5cm]{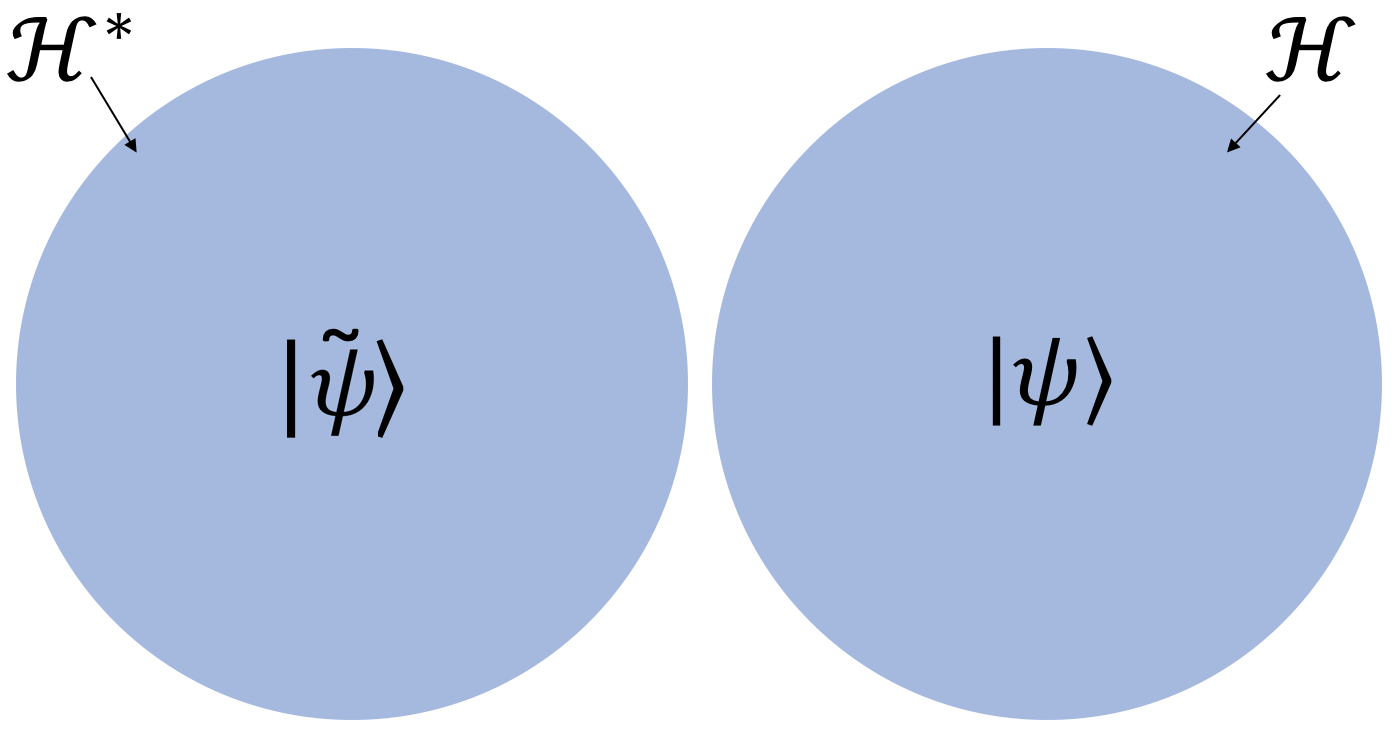}
\caption{A schematic diagram of a biorthogonal system. If a state $\ket{\psi}\in \mathcal{H}$ then it evolves with $H$ and its associated state $\ket{\widetilde{\psi}}\in\mathcal{H}^*$ evolves with $H^{\dagger}$.
\label{fig:biovenn}}
\end{figure}
\unskip

The time dependence of states that belong to the Hilbert space $\mathcal{H}$ can be calculated using the time-dependent Schr\"odinger equation, which implies that 
\begin{equation} \label{22}
|\psi(t)\rangle = e^{-\frac{i}{\hbar}Ht}|\psi(0)\rangle 
\end{equation} 
for a given initial state $|\psi(0)\rangle$. However, the associated states $|\widetilde{\psi}\rangle$, which are shown in Fig.~\ref{fig:biovenn} and belong to the Hilbert space $\mathcal{H}^*$, evolve with the Hermitian conjugate of $H$ such that
\begin{equation} \label{23}
|\widetilde{\psi}(t)\rangle = e^{-\frac{i}{\hbar}H^\dagger t}|\widetilde{\psi}(0)\rangle
\end{equation}
for a given initial associated state $|\widetilde{\psi}(0)\rangle$. In contrast to conventional Hermitian quantum physics, states do not all evolve according to the same Hamiltonian. Again, the two Hilbert spaces $\mathcal{H}$ and $\mathcal{H}^*$ differ solely by their inner products since both contain the same vector space. 

Looking at the basis states, the initial state $|\psi(0)\rangle = |\alpha_n\rangle$ evolves into the state 
\begin{equation}
|\psi(t)\rangle = e^{-\frac{i}{\hbar}E_n t}|\alpha_n \rangle 
\end{equation}
because the states $|\alpha_n\rangle$ are eigenstates of $H$ with eigenvalues $E_n$. Using Eqs.~(\ref{PH1}) and (\ref{beta1}) one can show that the state $|\beta_n\rangle$ is an eigenstate of $H^\dagger$ with the same eigenvalue, $E_n$, as its orthogonal partner $|\alpha_n\rangle$.  Therefore, the initial state $|\widetilde{\psi}(0)\rangle = |\beta_n\rangle$ evolves into
\begin{equation}
|\widetilde{\psi}(t)\rangle = e^{-\frac{i}{\hbar}E_n t}|\beta_n \rangle\,.
\end{equation}
Thus, $|\alpha_n\rangle$ and $|\beta_n\rangle$ evolve identically. Nevertheless, orthogonal states under the inner product of $\mathcal{H}$ are, in general, not orthogonal under the inner product of $\mathcal{H}^*$ and vice versa. Hence, evolving states unitarily requires that the states in $\mathcal{H}$ and in $\mathcal{H}^*$ experience different Hamiltonians.

\section{A local non-Hermitian description of the quantised EM field}

Writing the electric and magnetic field observables in terms of local ($x$-dependent) ladder operators as
\begin{eqnarray} \label{localfieldsA}
 {\bf E}(x)&=&\sum_{s,\lambda} \sqrt{\frac{\hbar c}{2\varepsilon A}} \, {A}_{s\lambda}(x) \,  {\bf e}_\lambda+ H.c. \, , \nonumber \\
 {\bf B}(x)&=&\sum_{s,\lambda} {s \over c} \, \sqrt{\frac{\hbar c}{2\varepsilon A}} \,{A}_{s\lambda}(x) \,  {\bf e}_x\times {\bf e}_\lambda + H.c.\, ,
\end{eqnarray}
the above expressions are consistent with the standard momentum-space field observables given in Eq.~(\ref{localfields}) when the annihilation operators $A_{s\lambda}(x)$ are given by
\begin{equation}\label{A tilde}
{A}_{s\lambda}(x) = \int_{0}^\infty {\rm d} k \, \sqrt{k \over 2 \pi} \, {\rm e}^{{\rm i} skx} \, a_{s\lambda}(k) \, .
\end{equation}
Using Eq.~(\ref{com1}), we find that the above annihilation operator commutes with itself for any $s, \lambda$ and $x$, as does its Hermitian conjugate. However, the commutator between annihilation and creation operators is non-zero and equals  
\begin{equation} \label{28}
	\left[{A}_{s\lambda}(x),{A}^{\dagger}_{s'\lambda'}(x') \right] =\delta_{ss'}\delta_{\lambda\lambda'}\frac{1}{2\pi}\int_0^{\infty}{\rm d}k\, k \, e^{isk(x-x')}\,.
\end{equation}
This commutator is not locally bosonic; that is, it is not proportional to $\delta(x-x')$. As such, the argument in Eq.~(\ref{6}) does not apply, and the single-excitation states ${A}^\dagger_{s\lambda}(x) |0 \rangle$ and ${A}^\dagger_{s\lambda}(x') |0 \rangle$ are not pairwise orthogonal. This means that we cannot interpret ${A}^\dagger_{s\lambda}(x)$ as the creation operator for a single excitation localised at $x$. 

\subsection{Orthogonal local field excitation states}

At this point, we will deviate from the standard description of quantum electrodynamics \cite{EJP} in order to produce annihilation operators for field excitations that are locally bosonic.  As in our earlier papers \cite{Jake, Daniel,Casimir}, we suppose that the allowed range of frequencies $k$ is extended to include all real values encompassing the negative frequencies as well as the positive ones, while Eqs.~(\ref{localfieldsA})-(\ref{28}) remain valid up to a change of the lower integral limit in Eqs.~(\ref{A tilde}) and (\ref{28}) to negative infinity. In addition, we replace $k$ in Eqs.~(\ref{A tilde}) and (\ref{28}) by $|k|$. In other words, we replace the operators ${A}_{s\lambda}(x)$ in Eq.~(\ref{A tilde}) by two new operators, namely
\begin{eqnarray}\label{A dagger bio 1}
A_{s\lambda}(x) &=& \int_{-\infty}^\infty {\rm d} k \, \sqrt{|k| \over 2 \pi} \, {\rm e}^{{\rm i} skx} \, a_{s\lambda}(k) \, , \nonumber \\
A_{s\lambda}^{bio}(x) &=& \int_{-\infty}^\infty {\rm d} k \, \sqrt{1 \over 2 \pi|k|} \, {\rm e}^{{\rm i} skx} \, a_{s\lambda}(k) \,.
\end{eqnarray}
When we calculate the commutator relations for these operators, with the help of Eq.~(\ref{com1}), we now find that
\begin{equation}\label{commutator1}
\left[A^{bio}_{s\lambda}(x), A^{\dagger}_{s'\lambda'}(x') \right]=\delta_{ss'}\delta_{\lambda\lambda'}\delta(x-x')\, ,
\end{equation}
which is different from Eq.~(\ref{28}). Therefore, $A^\dagger_{s\lambda}(x)$ can be considered to be a bosonic creation operator while $A^{bio}_{s\lambda}(x)$ is the corresponding annihilation operator. Clearly $A^{bio}_{s\lambda}(x)$ is not the Hermitian conjugate of $A^\dagger_{s\lambda}(x)$. Usually, this would be a problem; however, if the single-photon state $A^{bio\,\dagger}_{s\lambda}(x)|0\rangle$ is the associated state of $A^{\dagger}_{s\lambda}(x)|0\rangle$, we obtain a biorthogonal system in which the $A^{\dagger}_{s\lambda}(x)$ states are locally bosonic.

In other words, we need to alter the conventional inner product of quantum physics. In the following, we therefore identify an inner product such that the annihilation operator corresponding to the creation operator $A^\dagger_{s\lambda}(x)$ is indeed $A^{bio}_{s\lambda}(x)$. However, before proceeding, there are a couple of points that we must be aware of. Firstly, by taking the Hermitian conjugate of Eq.~\eqref{commutator1} we find that 
\begin{equation}\label{commutator2}
\left[A_{s\lambda}(x), A^{bio\,\dagger}_{s'\lambda'}(x') \right]=\delta_{ss'}\delta_{\lambda\lambda'}\delta(x-x')\,.
\end{equation}
Therefore the states generated by $A^{bio\,\dagger}_{s\lambda}(x)$ are locally bosonic when their annihilation operators are given by the $A_{s\lambda}(x)$ operators.  This means that we now have two distinct pairs of locally commuting Fock operators: we have the $\{A^{bio\,\dagger}_{s\lambda}(x), A_{s\lambda}(x)\}$ pair, and we also have the $\{A^{\dagger}_{s\lambda}(x), A^{bio}_{s\lambda}(x)\}$ pair. Because the EM field observables in Eq.~(\ref{localfieldsA}) are a linear sum of $A_{s\lambda}(x)$ and $A^{\dagger}_{s\lambda}(x)$, the field observables are expressed in terms of creation and annihilation operators that do not belong to the same pair of Fock operators. This must be taken into account when we construct the corresponding Hamiltonians $H$ and $H^\dagger$ of the quantised EM field in the position representation.

Secondly, we need an inner product under which the field excitations are locally bosonic. It is important that the introduction of such an inner product does not spoil the orthogonality of the monochromatic photons, thereby causing previously normalised states to be non-normalisable. Taking care that this is not the case and preserving the orthogonality of the monochromatic states is important because it allows us, for example, to construct coherent states of the EM field that oscillate like classical sinusoidal waves. In the following, we therefore adjust the standard inner product accordingly. To proceed we define a single local and a single bio-local excitation state, respectively, as
\begin{eqnarray}
\ket{1_{s\lambda}(x)} &=& A^{\dagger}_{s\lambda}(x)\ket{0}\,,
\nonumber \\
\ket{1_{s\lambda}(x)}^{bio} &=& A^{bio\,\dagger}_{s\lambda}(x)\ket{0}\, .
\end{eqnarray}
In order to utilise the bosonic commutation relations in Eqs.~\eqref{commutator1} and \eqref{commutator2}, when analysing the dynamics of expectation values we require that the above states be pairwise orthonormal. However, under the standard inner product
\begin{eqnarray}
\braket{\ket{1_{s\lambda}(x)},\ket{1_{s\lambda}(x')}} &\ne & \delta(x-x')\, , \nonumber \\
\braket{\ket{1_{s\lambda}(x)}^{bio},\ket{1_{s\lambda}(x')}^{bio}} &\ne & \delta(x-x')\,.
\end{eqnarray}
To address this, using the commutator in Eq.~(\ref{commutator1}) one can show that for every state $\ket{1_{s\lambda}(x)}$ there is a state $\ket{1_{s\lambda}(x')}^{bio}$ such that 
\begin{eqnarray}
\braket{\ket{1_{s\lambda}(x')}^{bio},\ket{1_{s\lambda}(x)}}= \delta(x-x')\,.
\end{eqnarray}
Therefore, it would be useful to identify the EM field in the following with a biorthogonal system where $\ket{1_{s\lambda}(x)}^{bio}$ is the associated state of $\ket{1_{s\lambda}(x)}$. Likewise, for every $\ket{1_{s\lambda}(x)}^{bio}$ there is a state $\ket{1_{s\lambda}(x')}$ such that 
\begin{eqnarray}
\braket{\ket{1_{s\lambda}(x')},\ket{1_{s\lambda}(x)}^{bio}}= \delta(x-x')\,.
\label{eq:orthogonalstates}
\end{eqnarray}

To provide a connection with the previous section, we may initially try to define appropriate $\eta$ and $\eta^{-1}$ operators. As shown in App.~\ref{app A}, these are given by \cite{Ali}
\begin{eqnarray}\label{eta and eta inverse}
\eta&=& \sum_{s, \lambda}\int_{- \infty}^\infty {\rm d} x \ket{1_{s\lambda}(x)}^{bio}\bra{1_{s\lambda}(x)}^{bio},
\nonumber \\
\eta^{-1}&=& \sum_{s, \lambda}\int_{- \infty}^\infty {\rm d} x \ket{1_{s\lambda}(x)}\bra{1_{s\lambda}(x)}\,.
\end{eqnarray}
With these operators, we can define an $\eta$ and an $\eta^{-1}$ inner product by 
\begin{eqnarray}\label{bioin1}
\braket{\ket{\psi},\ket{\phi}}^{\eta}&=&\bra{\phi}\eta\ket{\psi}\,,
\nonumber \\
\braket{\ket{\psi},\ket{\phi}}^{\eta^{-1}}&=&\bra{\phi}\eta^{-1}\ket{\psi}\,,
\end{eqnarray}
from which it follows that
\begin{eqnarray}
\braket{\ket{1_{s\lambda}(x')},\ket{1_{s\lambda}(x)}}^{\eta} &=& \braket{\ket{1_{s\lambda}(x')}^{bio},\ket{1_{s\lambda}(x)}^{bio}}^{\eta^{-1}} \notag \\
&=& \delta(x-x')\,.
\end{eqnarray}
Now the local and bio-local states are pairwise orthonormal with respect to the $\eta$ and the $\eta^{-1}$ inner product respectively.
Unfortunately, neither of the above inner products is satisfactory for our purposes because if we, for example, apply them to photon states and bio-local states we find that
\begin{eqnarray}
\braket{\ket{1_{s\lambda}(k')},\ket{1_{s\lambda}(k)}}^{\eta^{-1}}&\ne& \delta(k-k')\,,
\nonumber \\
\braket{\ket{1_{s\lambda}(x')}^{bio},\ket{1_{s\lambda}(x)}^{bio}}^{\eta}&\ne& \delta(x-x')\,.
\end{eqnarray}

\subsection{A generalised inner product}
\label{3IP}

What we want is an inner product where the local, bio-local and monochromatic photon states are all pairwise orthonormal. Therefore, we shall next describe a more suitable and general way of defining the biorthogonal conjugate of a given state vector. To achieve this, we first replace $\sqrt{|k|}$ in Eq.~\eqref{A dagger bio 1} with a general (real) function $f(k)$. This means we replace $A_{s \lambda}(x)$ and $A^{bio}_{s \lambda}(x)$ by the two operators
\begin{eqnarray}
	\label{B ops}
    A_{s \lambda}(x) &=& \int_{-\infty}^\infty {\rm d} k  \frac{f(k)}{\sqrt{2 \pi}} {\rm e}^{{\rm i} skx} \, {a}_{s \lambda} (k)\, ,
    \nonumber \\
    A^{bio}_{s \lambda}(x) &=& \int_{-\infty}^\infty {\rm d} k  \frac{1}{f(k)\sqrt{2 \pi}} {\rm e}^{{\rm i} skx} \, {a}_{s \lambda} (k)\,.
\end{eqnarray}
Here $f(k)$ should be chosen such that
\begin{equation}
{a}_{s \lambda}^\dagger(k) = \int_{-\infty}^\infty {\rm d} x \, \frac{1}{f(k)\sqrt{2 \pi}} \, e^{- iskx} \, A^\dagger_{s \lambda} (x) \, ,
\end{equation}
since this gives the correct commutation relation, leading to Eq.~(\ref{eq:orthogonalstates}). In the following we refer to $f(k)$, similarly to Ref.~\cite{Raymer}, as the Fourier weight function and demand that the $A^\dagger_{s\lambda}(x)$ operators generate local excitations. 

Consider now a state $|\psi\rangle$ where
\begin{eqnarray}
|\psi\rangle &=& \sum_{s, \lambda}\int_{-\infty}^{\infty}\text{d}x\; \psi_{s\lambda}(x)A^\dagger_{s\lambda}(x)|0\rangle\,, \nonumber \\
|\psi\rangle^{bio} &=& \sum_{s, \lambda}\int_{-\infty}^{\infty}\text{d}x\;\psi_{s\lambda}(x)A^{bio\,\dagger}_{s\lambda}(x)|0\rangle\,.
\end{eqnarray} 
Here $|\psi\rangle^{bio}$ is the associated state of $|\psi\rangle$. Let $S$ be an operator that inverts any Fourier weight terms contained within any state it acts upon. One can then see that $S$ maps $|\psi\rangle$ to an associated state $|\psi\rangle^{bio} = S(|\psi\rangle)$.  $S$ is also its own inverse since $S (|\psi\rangle^{bio}) = |\psi\rangle$. For a momentum state, i.e.~a state with
\begin{equation}
|\psi\rangle = \sum_{s, \lambda}\int_{-\infty}^{\infty}\text{d}k\; \psi_{s\lambda}(k) a^\dagger_{s\lambda}(k)|0\rangle\,,
\end{equation}
we have $S(|\psi\rangle) = |\psi\rangle$, since
\begin{eqnarray}
S\left(a_{s\lambda}^{\dagger}(k)\ket{0}\right) &=& S\left(\int_{-\infty}^\infty {\rm d} x \, {f(k) \over \sqrt{2 \pi}} \, {\rm e}^{ i skx} \, A^{\dagger\,bio}_{s\lambda}(x)\ket{0}\right) 
\nonumber \\
&=& \int_{-\infty}^\infty {\rm d} x \, \frac{1}{f(k)\sqrt{2 \pi}} \, {\rm e}^{ i skx} \, A^{\dagger}_{s\lambda}(x)\ket{0}
\nonumber \\
&=&a_{s\lambda}^{\dagger}(k)\ket{0}\,.
\end{eqnarray}
In momentum space, all states are equal to their associated states.  We therefore refer to them as photonic states.

For a given $f(k)$, let us define an inner product such that
\begin{eqnarray}
\braket{\ket{\psi},\ket{\phi}}^{bio}=S\left(\bra{\phi}\right)\ket{\psi}\,.
\end{eqnarray}
Under this inner product, 
\begin{eqnarray}
\braket{\ket{1_{s\lambda}(k')},\ket{1_{s\lambda}(k)}}^{bio}&=& \delta(k-k')\,,\nonumber \\
\braket{A^\dagger_{s \lambda} (x)\ket{0},A^\dagger_{s \lambda} (x)\ket{0}}^{bio}&=& \delta(x-x')\,,\nonumber \\
\braket{A^{\dagger\,bio}_{s \lambda} (x)\ket{0}, A^{\dagger\,bio}_{s \lambda} (x)\ket{0}}^{bio}&=&\delta(x-x')\,.
\end{eqnarray}
Using the above generalised inner product, the photon as well as the local and the bio-local states all form pairwise orthogonal sets.  We therefore use this generalised inner product in the following to model the quantised EM field.  The biorthogonal conjugate, or bio-conjugate, $O^{bio}$ of an operator $O$ can now be defined as the operator that satisfies
\begin{equation}
\langle |\psi\rangle, O^{bio\, \dagger}|\phi\rangle \rangle^{bio} = \langle O|\psi\rangle, |\phi\rangle \rangle^{bio}\,.
\end{equation}
To calculate the biorthogonal conjugate $O^{bio}$ of an operator $O$ that depends on $f(k)$, we must replace $f(k)$ with its reciprocal, $1/f(k)$.

If we are aiming for a description of the quantised EM field in which the local electric and magnetic field operators $ {\bf E}(x)$ and $ {\bf B}(x)$ each obey bosonic commutator relations, the Fourier weight function of the operators $A_{s \lambda}(x)$ and $A^{bio}_{s \lambda}(x)$ in Eq.~(\ref{B ops}) of field excitations needs to be
\begin{equation}
f(k) = \sqrt{|k|} 
\end{equation}
which again turns these operators into the $A_{s \lambda}(x)$ and $A^{bio}_{s \lambda}(x)$ operators in Eq.~(\ref{A dagger bio 1}). However, how we use these operators has now changed. States with the above weight function are localised in the sense that the corresponding field excitations generate local electric and magnetic field expectation values. Later on in Section \ref{sec4}, we will have a closer look at alternative definitions of local EM field annihilation and creation operators which are bosonic with respect to the conventional inner product of quantum physics and will pay more attention to the physical interpretations of the above operators.

\subsection{Time evolution in the biorthogonal representation of the EM field}

As shown in our earlier work \cite{Jake,Daniel,Casimir}, in this new biorthogonal description of the EM field, the free-space Hamiltonian $H$ that generates the dynamics of light is 
\begin{equation}\label{H free all wn}
H =\sum_{s,\lambda}\int_{-\infty}^{\infty}{\rm d}k\, \hbar c k\,a^{\dagger}_{s\lambda}(k)a_{s\lambda}(k) \, .
\end{equation}
Notice that this Hamiltonian no longer coincides with the energy observable $H_{eng}$ of the EM field. For example, despite the $a^{\dagger}_{s\lambda}(-k)\ket{0}$ being a negative-frequency state, i.e. an eigenstate of the Hamiltonian $H$ with a negative eigenvalue, it still has a well-defined positive energy expectation value. 
This is so because when we substitute the EM field observables into the classical energy observable in Eq.~\eqref{classical energy}, we obtain the positive operator \cite{Jake, Daniel}
\begin{eqnarray}
H_{eng}=\sum_{s,\lambda}\int_{-\infty}^{\infty}{\rm d}k\,\frac{\hbar c |k|}{2} \left(a_{s\lambda}(k) + H.c.\right)^2 \,.
\end{eqnarray}
When we restrict the Hilbert space of the EM field again to positive-frequency states, $H$ and $H_{eng}$ coincide perfectly, as they do in the standard description of the EM field \cite{EJP}. Here the negative-frequency photons have been added, since they ensure for example that wave packets of any shape can travel at the speed of light in one direction, i.e.~without dispersion.

However, before we can study the dynamics of electric and magnetic field expectation values, we must first examine how the dynamics of this system differs from both a conventional quantum system and a typical biorthogonal system. To ensure that our time-evolution operators $U(t)$ are unitary under the generalised inner product, we require the inner product between states in the Schr\"odinger picture to be constant in time. Given two states $\ket{\psi_0}$ and $\ket{\phi_0}$ at a time $t=0$, we require that
\begin{equation}
\braket{U(t)\ket{\psi_0},U(t)\ket{\phi_0}}^{bio} = \braket{\ket{\psi_0},\ket{\phi_0}}^{bio}
\end{equation}
which implies that $U^{bio\,\dagger}(t) U(t)={\rm I_{\, d}}$ and is true when
\begin{equation}\label{bioHermitian}
H^{bio\,\dagger}=H\, .
\end{equation}
In the following, we only consider Hamiltonians that satisfy this relation and refer to them as bio-Hermitian. Taking the Hermitian conjugate of each side of the above equation gives us $H^{bio}=H^\dagger$. Hence the bio-conjugate of a bio-Hermitian Hamiltonian is equal to its Hermitian conjugate and $H^{bio}$ and $H^{\dagger}$ can be used interchangeably. 

Fortunately, the condition in Eq.~(\ref{bioHermitian}) holds for the field Hamiltonian $H$ in Eq. (\ref{H free all wn}) which generates the dynamics of free photons. This is not surprising, since local and non-local photons all propagate alike. Because wave packets of any shape all propagate at the same speed $c$, the Hamiltonian does not depend on the particular choice of $f (k)$ that defines a local excitation. For a proof that $H$ equals its biorthogonal conjugate, see App.~\ref{app B}. 

\subsubsection{The dynamics of states}

\begin{figure}[t]
\centering
\includegraphics[width=5cm]{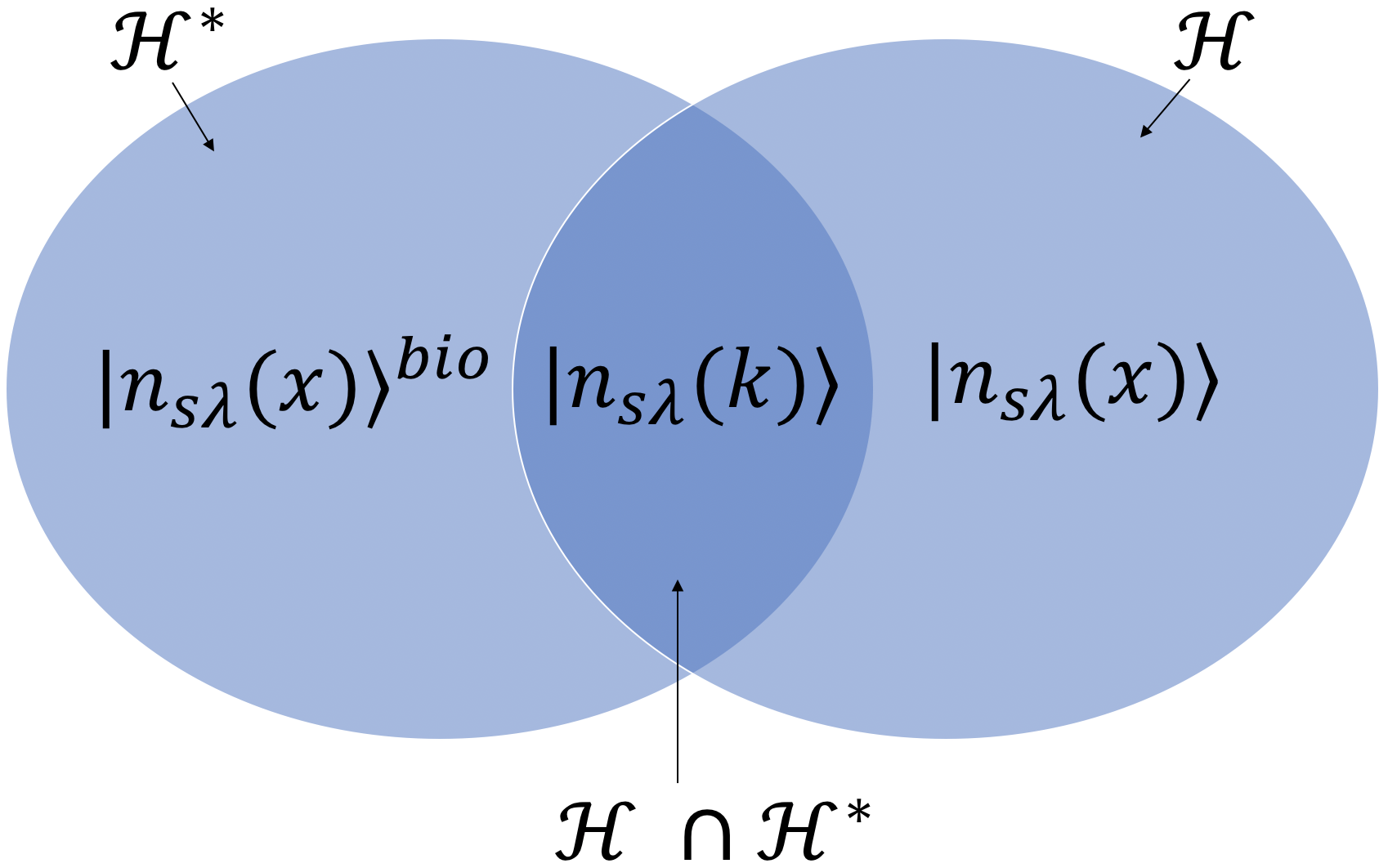}
\caption{A schematic diagram of our EM system. In contrast to the previous figure, the intersection between the Hilbert space of states and its bio-conjugate space is non-empty. Shown above are the number states of $\mathcal{H}^{*}$, $\mathcal{H}\cap\mathcal{H}^{*}$ and $\mathcal{H}$. Those states contained in $\mathcal{H}\cap\mathcal{H}^*$ are normalisable under either inner product.
\label{fig:blipvenn}}
\end{figure}
\unskip 

According to biorthogonal quantum physics, the EM field states in $\mathcal{H}$ evolve with $H$ and states in $\mathcal{H}^*$ evolve with $H^\dagger = H^{bio}$ \cite{Brody}. The general solutions of the corresponding Schr\"odinger equations can be found in Eqs.~(\ref{22}) and (\ref{23}). Next we therefore need to decide whether a state belongs to $\mathcal{H}$ or to $\mathcal{H}^*$. Before we do, however, we point out that $\mathcal{H}$ and $\mathcal{H}^*$ are both equipped with the same inner product and so can be thought of as part of a larger Hilbert space $\mathcal{H}\cup\mathcal{H}^*$. The key distinction between the two Hilbert space is the dynamics of their states.

In the following, $\mathcal{H}\setminus\mathcal{H}^{*}$ denotes the space spanned by the Fock states that are generated by acting the local creation operators $A^{\dagger}_{s\lambda}(x)$ in Eq.~(\ref{A dagger bio 1}) on the vacuum state. Similarly, $\mathcal{H}^{*}\setminus\mathcal{H}$ is spanned by the Fock states that are generated by acting the bio-local creation operators $A^{\dagger\,bio}_{s\lambda}(x)$ in Eq.~(\ref{A dagger bio 1}) on the vacuum state. Clearly, for every state in $\mathcal{H}$ its bio-conjugate state is in $\mathcal{H}^*$ and vice versa. Lastly, $\mathcal{H}\cap\mathcal{H}^{*}$ contains the Fock space spanned by the Fock states that are generated by acting monochromatic photon creation operators, $a^{\dagger}_{s\lambda}(k)$, on the vacuum state, since any monochromatic photon Fock state is its own bio-conjugate (see Fig.~\ref{fig:blipvenn}). An immediate question that arises, then, is whether it matters if a photonic state is evolved using $H$ or $H^{bio}$. The answer is {\em yes}, but to understand why this is the case, we need to look at how operators evolve in time.

\subsubsection{The dynamics of operators}

To identify if a state evolves according to $H$ or $H^{bio}$ in the Schr\"odinger picture, we look at whether the state is contained in $\cal{H}$ or $\cal{H}^{*}$. Similarly, to identify if an operator evolves using $H$ or $H^{bio}$ in the Heisenberg picture, we look at whether it acts on $\mathcal{H}$ or $\mathcal{H}^{*}$. This identification is perhaps easier to see in our system where we use Fock state operators to construct our operators and states. For example, because $A^{\dagger}_{s\lambda}(x)\ket{0}$ evolves using $H$ in the Schr\"odinger picture, both it and its corresponding annihilation operator, $A^{bio}_{s\lambda}(x)$, should evolve with respect to $H$ in the Heisenberg picture. Likewise, because $A^{\dagger bio}_{s\lambda}(x)\ket{0}$ evolves using $H^{bio}$ in the Schr\"odinger picture, then both it and its corresponding annihilation operator should evolve with respect to $H^{bio}$ in the Heisenberg picture. 

For example, suppose an operator $ {B}$ is of the form $ {B}=\sum_i {B}_i$, where for each $i$, $B_i$ is a map from either $\mathcal{H} \rightarrow\mathcal{H}$ or from $\mathcal{H}^{*} \rightarrow\mathcal{H}^{*}$.  If ${B_i}$ maps states from $\mathcal{H}$ to $\mathcal{H}$, we use $ {H}$ to generate its dynamics in the Heisenberg equation and
\begin{equation}\label{Hequation}
\frac{{\rm d}}{{\rm d}t} {B}_i(t) =-\frac{i}{\hbar}\left[B_i(t),H\right]
\end{equation}
with $B_i(t) = U^{\dagger bio}(t)B_iU(t)$.  Conversely, if it maps states from $\mathcal{H}^{*}$ to $\mathcal{H}^{*}$ we use $H^{bio}$ to generate its dynamics in the Heisenberg equation
\begin{equation}\label{Hbioequation}
\frac{{\rm d}}{{\rm d}t} {B}_i(t) =-\frac{i}{\hbar}\left[B_i(t),H^{bio}\right]
\end{equation}
with $B_i(t) = U^{\dagger}(t)B_iU^{bio}(t)$. This means that the local Fock operators, ${A}^{bio}_{s\lambda}(x)$ and ${A}^\dagger_{s\lambda}(x)$, evolve according to Eq.~(\ref{Hequation}), and the bio local Fock operators, ${A}_{s\lambda}(x)$ and ${A}^{\dagger bio}_{s\lambda}(x)$, evolve according to Eq.~(\ref{Hbioequation}). If $B_i=B^{bio}_i$, such as is the case with the $a_{s\lambda}(k)$ and $a_{s\lambda}^{\dagger}(k)$ operators, then either $H$ or $H^{bio}$ can be used to generate their dynamics.  

\subsubsection{The dynamics of expectation values}

When calculating the time-dependent expectation values of an operator, it should not matter whether the expectation value is calculated in the Schr\"odinger picture or in the Heisenberg picture. The same applies to the biorthogonal system that we consider here. For example, if we have an operator $B:\mathcal{H} \to \mathcal{H}$ and a state $\ket{\psi}\in \mathcal{H}$ then the corresponding expectation value is 
\begin{eqnarray}
\langle B\ket{\psi(t)}, \ket{\psi(t)} \rangle^{bio} &=& ^{bio}\langle\psi(t)|B|\psi(t)\rangle\nonumber\\
&=& ^{bio}\langle \psi|U^{\dagger bio}(t) B U(t) |\psi \rangle \nonumber\\
&=& ^{bio}\langle \psi |B(t)|\psi\rangle \nonumber \\
&=&\langle B(t)\ket{\psi}, \ket{\psi} \rangle^{bio}\,,
\end{eqnarray}
where $B(t)$ satisfies Eq.~(\ref{Hequation}). Similarly one can show that if we have an operator $B:\mathcal{H}^* \to \mathcal{H}^*$ and a state $\ket{\phi}\in \mathcal{H}^*$ then
\begin{eqnarray}
\langle B\ket{\phi(t)}, \ket{\phi(t)} \rangle^{bio} &=& ^{bio}\langle\phi(t)|B|\phi(t)\rangle\nonumber\\
&=& ^{bio}\langle \phi|U^{\dagger}(t) B U^{bio}(t) |\phi \rangle \nonumber\\
&=&^{bio}\langle \phi |B(t)|\phi\rangle \nonumber \\
&=&\langle B(t)\ket{\phi}, \ket{\phi} \rangle^{bio}\,, 
\end{eqnarray}
where $B(t)$ satisfies Eq.~(\ref{Hbioequation}). Therefore, expectation values can only be valid if the Schr\"odinger and Heisenberg pictures agree.

If we have an operator that satisfies $B=B^{bio}$ and a photonic state with $\ket{\psi}^{bio}=\ket{\psi}$, using either $H$ or $H^{bio}$ leads to the same real expectation value with respect to the generalised inner product so long as $B$ is bio-Hermitian.  This is so because
\begin{eqnarray}
\braket{BU(t)\ket{\psi},U(t)\ket{\psi}}^{bio}&=&\bra{\psi}{U}^{ \dagger\, bio}(t)B{U}(t)\ket{\psi}
\nonumber \\
&=&(\bra{\psi}{U}^{\dagger\, bio}(t)B{U}(t)\ket{\psi})^*
\nonumber \\
&=&\bra{\psi}{U}^{\dagger}(t)B{U}^{bio}(t)\ket{\psi}
\nonumber \\
&=&\braket{{BU^{bio}(t)\ket{\psi}},U^{bio}(t)\ket{\psi}}^{bio} \, . \notag \\
\end{eqnarray}
In the second line we have used the property that the expectation value is real and is therefore equal to its complex conjugate, which is denoted by the asterisk. In the third line we have used the property that $B$ is Hermitian, since it is both bio-Hermitian and its own bio-conjugate.  

There are certain operators, however, such as the electric and magnetic field observables, that contain operators acting on both $\mathcal{H}$ and $\mathcal{H}^*$. As can be seen from Eqs.~(\ref{localfieldsA}) and (\ref{A dagger bio 1}), we need to use both $H$ and $H^{bio}$ to determine the time evolution of the expectation value of such an operator.  As an example, suppose we have the Hermitian, but not bio-Hermitian, operator ${A}={A}_{s\lambda}(x)+{A}^\dagger_{s\lambda}(x)$. This operator evolves in the Heisenberg picture as 
\begin{equation}
A(t)=U^\dagger(t){A}_{s\lambda}(x)U^{bio}(t) + U^{\dagger bio}(t){A}^\dagger_{s\lambda}(x){U}(t)
\end{equation}
and so remains Hermitian as time passes, since $A(t)=A^{\dagger}(t)$. The corresponding expectation value of this operator with respect to a photonic state $\ket{\psi}=\ket{\psi}^{bio}$ is the real quantity
\begin{eqnarray}
\braket{A(t)\ket{\psi},\ket{\psi}}^{bio}&=&
\braket{A_{s\lambda}(x)U^{bio}(t)\ket{\psi},U^{bio}(t)\ket{\psi}}^{bio} \notag \\
&& +\braket{A_{s\lambda}^{\dagger}(x)U(t)\ket{\psi},U(t)\ket{\psi}}^{bio} \nonumber \\
&=&\braket{\psi|A(t)|\psi}\,,
\label{eq:matrixelement}
\end{eqnarray}
and is a valid expectation value because photonic states evolve in the same way when using either $H$ or $H^{bio}$. The last line is simply the expectation value of a Hermitian operator in the standard inner product, which is real. However, within this expectation value in the Schr\"odinger picture, the photonic state still has to evolve with the form of the Hamiltonian such that the expectation value agrees with the Heisenberg-picture expectation value. This is why the states in the first line above evolve with $H^{bio}$ but the states in the second line evolve with $H$. Again, just because a photonic state can evolve with either $H$ or $H^{bio}$ when considered by itself, it does not mean that within an expectation value it can evolve with either. It has to evolve such that it agrees with the Heisenberg picture; otherwise, the expectation value is not valid.

However, we cannot calculate the time-dependent expectation value of $A$ with respect to non-photonic states, because these states evolve using only $H$ or only $H^{bio}$. The corresponding expectation values in the Heisenberg and Schr\"odinger pictures would therefore not be in agreement.

\section{The connection with Hermitian descriptions} \label{sec4}

\subsection{Local bosonic excitations}

\begin{figure*}[t]
\includegraphics[width=0.7 \textwidth]{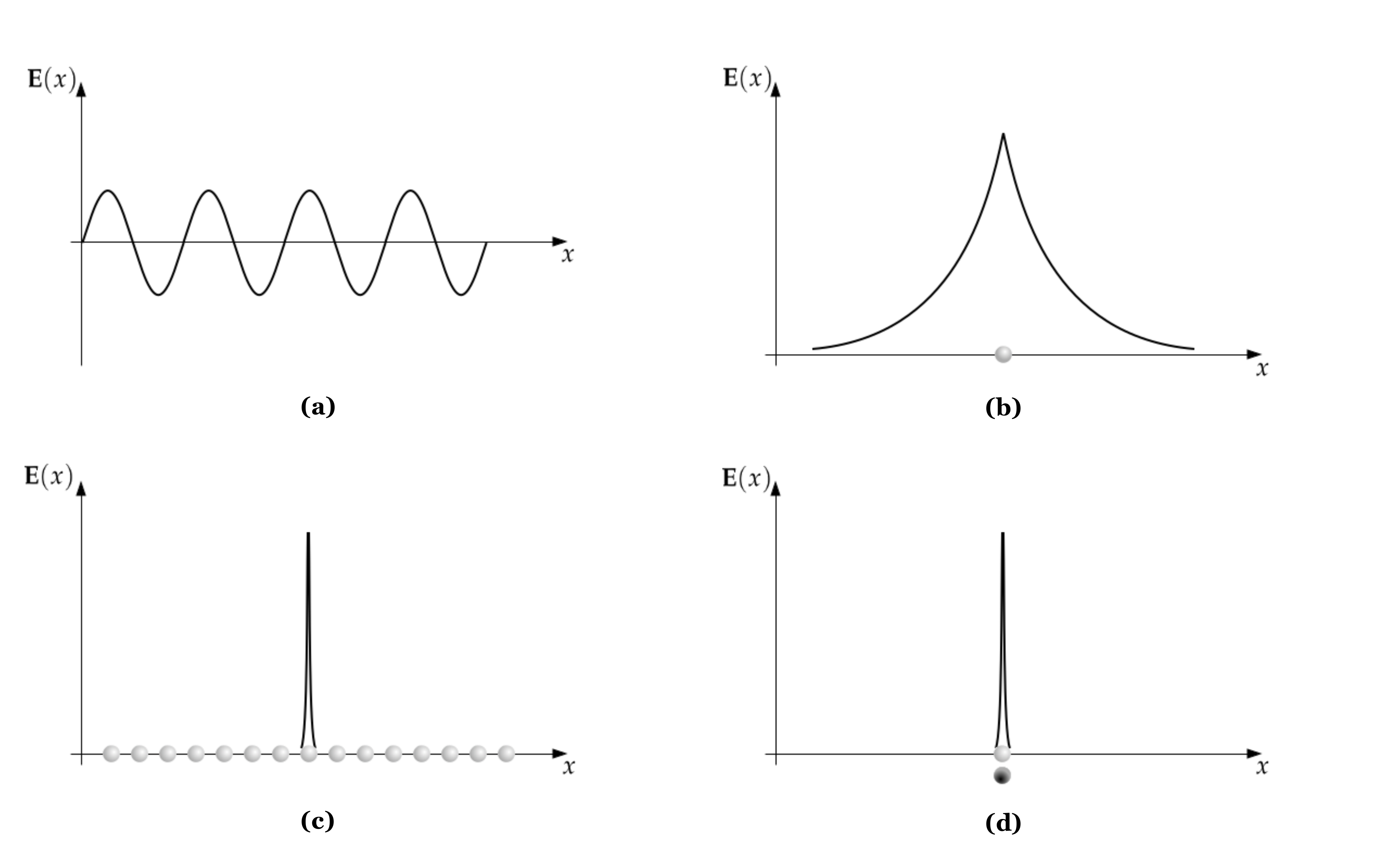}
\caption{Comparing the locality of different descriptions. In (\textbf{a}), momentum states are completely delocalised, with local excitations leading to superluminal propagation. In (\textbf{b}), a single blip allows for a local excitation that travels at $c$. However, if we choose the excitation to be local, then the fields associated with the excitation are not completely localised. In order to localise the field, we must superpose a non-local set of blips, as in (\textbf{c}). Since each blip travels at $c$, the resulting field configuration also travels travels at $c$. In (\textbf{d}), we localise both the excitation and the field by introducing a non-standard scalar product. This requires us to keep track of both the state and the conjugate state, since these in principle now evolve differently.}
\label{fig:comparison}
\end{figure*}

The standard approach to EM field quantisation is to work with the momentum eigenstates. Due to the uncertainty principle, these states are necessarily completely delocalised and can be thought of as waves that fill the volume under consideration, as in Fig.~\ref{fig:comparison}(a). Attempts to define a local excitation in this approach run into problems. While it is possible to produce an instantaneously localised field configuration with a suitable Fourier sum at a time $t=0$, this configuration spreads infinitely quickly for all times $t\neq 0$. That is, attempts to produce local excitations in the standard approach lead to apparent superluminal propagation. However, in a recent paper \cite{Daniel} we showed that it is possible to introduce local excitations by allowing for both positive-frequency and negative-frequency states, with
\begin{equation} \label{nokblips}
a_{s\lambda}(x) = {1 \over\sqrt{2 \pi}} \int_{-\infty}^\infty {\rm d} k  \, {\rm e}^{{\rm i} skx} \, a_{s\lambda}(k)\,.
\end{equation}
These $a_{s\lambda}(x)$ operators satisfy the commutation relation
\begin{equation}\label{blip commutator}
[a_{s\lambda}(x), a^{\dagger}_{s'\lambda'}(x')]=\delta_{ss'}\delta_{\lambda\lambda'}\delta(x-x')
\end{equation}
meaning they are locally bosonic under the conventional inner product. For this reason, we named these excitations blips (boson localised in position). The defining equation of motion for these blip operators guarantees propagation at the speed of light. 

We have successfully utilised these blips to quantise the EM field in position space \cite{Daniel} and construct locally acting mirror Hamiltonians \cite{Jake}. For the latter, we showed that it is possible to construct a mirror Hamiltonian that reproduces the classical mirror image effect for the local operators $a_{s\lambda}(x)$. This Hamiltonian was a significant find because, until that point, no locally acting mirror Hamiltonian for the EM field had been derived, and there is much interest in this topic in the literature \cite{Carniglia, Nick, Ben, Agarwal, Creatore}. Thus, there are certain situations in which blips provide a more physically intuitive description for modelling light-matter interactions compared to monochromatic photons.

Blips act on a Hilbert space of the same dimensions as that on which the operators $A_{s\lambda}(x)$ act.  In both cases, creation operators generate excitations that are characterised by a position $x$, a direction of propagation $s$ and a polarisation $\lambda$.  Furthermore, because the pairs of Fock operators $\{A_{s\lambda}^{bio}(x), A^\dagger_{s\lambda}(x)\}$ and $\{a_{s\lambda}(x), a^\dagger_{s\lambda}(x)\}$ have the same commutation relations there is a one-to-one correspondence between the two.  Therefore either can be used as a representation of the local excitations of the EM field.  

In the context of the generalised inner product we have utilised so far, this correspondence can be emphasised by pointing out that the conventional inner product can be thought of as a specific example of the generalised inner product in which blips are the bio-conjugate of themselves.  As a consequence of this, for any operator $O$ we find that $O^{bio} = O$ in both the position and momentum representations.  Furthermore, this means that all states evolve alike according to a single time-evolution operator $U(t) = U^{bio}(t)$.  This implies, therefore, that if we view the $A^\dagger_{s\lambda}(x)$ states as being localised and evolving according to a bio-Hermitian Hamiltonian, then we can also view an $a^\dagger_{s\lambda}(x)$ as being localised and reproduce the exact same dynamics using a Hermitian rather than bio-Hermitian Hamiltonian.  Since only a subset of Hermitian operators are also bio-Hermitian, however, these two Hamiltonians are in general not the same. 

In section \ref{3IP}, we noted that the dynamical Hamiltonian is both Hermitian and bio-Hermitian, even when using the generalised inner product.  That is to say that $H^{bio\,\dagger} = H = H^\dagger$; see App.~\ref{app B} for more information.  Such a Hamiltonian generates unitary dynamics under the conventional and generalised inner products, and imparts the exact same dynamics on both the $A^\dagger_{s\lambda}(x)$ and $a^\dagger_{s\lambda}(x)$ operators:
\begin{eqnarray}
\label{unitary1}
U^{\dagger}(t)a_{s\lambda}(x)U(t)=a_{s\lambda}(x-sct)\,.
\nonumber \\
U^{\dagger\, bio}(t)A_{s\lambda}(x)U(t)=A_{s\lambda}(x-sct)\,.
\end{eqnarray}
Of course here $U^{\dagger\, bio}(t)=U^{\dagger}(t)$, but for a more general Hamiltonian this may not be true, and we would no longer see an equivalence in the dynamics of the fields and the blips when we have only a single Hamiltonian as we do above.  For more information see App.~\ref{app C}. From the above equations, both the blips and local field modes therefore propagate at the speed of light along the $x$-axis in the direction specified by the parameter $s$.

Although both of these excitations can be interpreted as local excitations under a suitable inner product, and identical unitary dynamics can be generated for both states, there are pros and cons to both descriptions. In the blip description, because all states are the bio-conjugate of themselves there is only one pair of locally bosonic Fock operators, as opposed to two pairs in the generalised scheme.  This is because there is a single representation for a localised state.  Furthermore, again because all states are equal to their bio-conjugate states, all states evolve according to the same Hamiltonian, meaning that the Schr\"odinger and Heisenberg pictures are always in agreement.

\subsection{Field observables}

Using the blip operators as defined above, we previously showed in \cite{Daniel} that the EM field observables, Eq.~(\ref{localfields}), can be expressed in the form
\begin{eqnarray} \label{localfields2}
 {\bf E}(x)&=&\sum_{s,\lambda} \sqrt{\frac{\hbar c}{2\varepsilon A}} \, \mathcal{R}({a}_{s\lambda}(x)) \,  {\bf e}_\lambda+ H.c. \, , \nonumber \\
 {\bf B}(x)&=&\sum_{s,\lambda} {s \over c} \, \sqrt{\frac{\hbar c}{2\varepsilon A}} \, \mathcal{R}({a}_{s\lambda}(x)) \,  {\bf e}_x\times {\bf e}_\lambda + H.c. \, , ~~
\end{eqnarray}
where
and $\mathcal{R}$ is a superoperator such that
\begin{equation}
\mathcal{R}(a_{s\lambda}(k))=\sqrt{|k|}a_{s\lambda}(k)\,.
\end{equation}

The nature of the superoperator $\mathcal{R}$ is to smear out the field around the blip. That is, while the blip exists at a single point in space, the field associated with that blip is spread out, with a maximum expectation value at the location of the blip, as in Fig.~\ref{fig:comparison}(b). As a result, the commutation relations for the fields are not standard bosonic relations and have a non-zero overlap at non-zero displacements. We can interpret this in one of two ways. Firstly, we can consider the field fundamental, and think of the blips as the mean position of field excitations. Alternatively, we can consider the blips the fundamental entities and think of them as ``carrying'' around a non-local field. It is possible to localise the fields in the blip approach, but this requires the introduction of a non-local set of blips, as in Fig.~\ref{fig:comparison}(c). If we were to insist that both the excitations and their associated fields simultaneously obey bosonic relations, we require that the field observables are directly proportional to the ladder operators. This can only be achieved with a non-standard scalar product. Here we see the difference in approach. The standard approach would be to attempt to localise the fields, which leads to problems. The blip approach localises the underlying ``field carriers'', which requires that the fields themselves become delocalised. Finally, the non-Hermitian approach localises both the fields and the underlying excitations, at the expense of the standard inner product. In this approach, we must consider both the state and the conjugate state, as in Fig.~\ref{fig:comparison}(d), since their evolution is distinct. 

On first inspection, this seems to suggest that the non-Hermitian formalism is superior. However, it should be noted that each approach introduces additional subtleties, and that both contain aspects of non-locality. For example, in the case of an optical cavity, the non-local field associated with a blip has a non-trivial interaction with the cavity walls even when the blip is not at the boundary. Indeed, we have shown in Ref.~\cite{Casimir} that it is this non-local interaction that leads to the Casimir effect in this formalism. If we were to use the non-Hermitian formalism to model the same optical cavity, we would first need to find the appropriate scalar product for the cavity. In general, the appropriate scalar product would be dependent on the boundary conditions of the system under consideration. That is, while individual calculations in the non-Hermitian approach may appear truly localised, non-local effects have already been introduced in modelling the particular situation. 

The reality is that the two approaches are equivalent, with each a re-framing of the other. Since only matrix elements are measurable, we can use the differing inner products to transform from one formalism to the other, as we saw in Eq.~\eqref{eq:matrixelement}. Thus, we really have two parametrisations of the same formalism. In one, we can simplify calculations by using the fact that all commutation relations are bosonic, and in the other we can simplify calculations by using the fact that the scalar product is the standard product and states are their own conjugates. Which approach works best will undoubtedly depend on the particular scenario to be modelled.

\section{Conclusions}	

This paper has shown how to model the EM field using a non-Hermitian approach that utilises biorthogonal physics and negative wavenumbers. A key finding was that we represented the EM field as a biorthogonal system with a non-zero intersection between its Hilbert space and bio-conjugate Hilbert space. Consequently, the EM field observables at a point $x$ were a linear sum of a Fock bosonic creation operator, $A^{\dagger}_{s\lambda}(x)$, and a Fock bosonic annihilation operator, $A_{s\lambda}(x)$, that did not belong to the same creation-annihilation pair. This had implications for calculations; for example, $A^{\dagger}_{s\lambda}(x)$ evolved using $H$ in the Heisenberg equation whereas $A_{s\lambda}(x)$ evolved using $H^{\dagger}$, where $H$ need not be equal to $H^{\dagger}$. In contrast, for an expectation value with a photonic operator and with a photonic state, either $H$ or $H^{\dagger}$ can generate dynamics because they lead to the same expectation value. Within this system, we used an inner product under which the EM field observables were not Hermitian. However, for this inner product, we showed how these observables still gave real expectation values for certain states, including photon coherent states and normalised monochromatic photon states.

To the best of our knowledge, there is nothing in the literature concerning mathematical structures of the form in Fig.~\ref{fig:blipvenn}, where we have a biorthogonal system with a non-zero intersection between its Hilbert space and bio-conjugate Hilbert space. In this paper we showed how to use such a biorthogonal system to model the EM field. This paper is not, however, a paper on the general properties of a biorthogonal system with a non-zero intersection between its Hilbert and bio-conjugate Hilbert spaces. We leave this to the mathematical physicists to explore. Nevertheless, we expect authors to find interest in our work: for example, in the quantum optics community where spatial properties of light are concerned \cite{Eberly, Keller1, Keller2}, as this will help aid our understanding in light-matter interactions \cite{Forn-Diaz, Engel}; in quantum information, where researchers are increasingly utilising various modes of the EM field \cite{Brecht, Slussarenko, Knill}; in the non-Hermitian community, where researchers are applying non-Hermitian and, in particular, biorthogonal quantum mechanics to physical systems \cite{Kunst, Brody, Hawton_2017, ElGanainy, ElGanainy2}.

\noindent
{\em Acknowledgement.} J.S. and A.B. acknowledge financial support from the UK Engineering and Physical Sciences Research Council (EPSRC) through the Oxford Quantum Technology Hub NQIT (Grant Nr. EP/M013243/1). Moreover D.H. acknowledges an EPSRC PhD studentship (Award Ref. Nr. 2130171). Statement of compliance with EPSRC policy framework on research data: This publication is theoretical work that does not require supporting research data.


\appendix
\section{Calculation of $\eta\eta^{-1}$}\label{app A}

By using the definitions of $\eta$ and $\eta^{-1}$ given in Eq.~\eqref{eta and eta inverse} we can calculate
\begin{widetext}
\begin{eqnarray}
\eta\eta^{-1}&=&\sum_{s, \lambda}\int_{- \infty}^\infty {\rm d} x \ket{1_{s\lambda}(x)}^{bio}\bra{1_{s\lambda}(x)}^{bio} \sum_{s', \lambda'}\int_{- \infty}^\infty {\rm d} x' \ket{1_{s'\lambda'}(x')}\bra{1_{s'\lambda'}(x')}
\nonumber \\
&=&\sum_{s, \lambda}\sum_{s', \lambda'}\int_{- \infty}^\infty {\rm d} x\int_{- \infty}^\infty {\rm d} x'\delta_{ss'}\delta_{\lambda\lambda'}\delta(x-x') \ket{1_{s\lambda}(x)}^{bio}\bra{1_{s'\lambda'}(x')}
\nonumber \\
&=&\sum_{s, \lambda}\int_{- \infty}^\infty {\rm d} x\,A^{\dagger\,bio}_{s\lambda}(x)\ket{0}\bra{0}A_{s\lambda}(x)
\nonumber \\
&=&\sum_{s, \lambda}\int_{- \infty}^\infty {\rm d} k\int_{- \infty}^\infty {\rm d} k'\sqrt{\frac{|k'|}{|k|}} \left(\frac{1}{2\pi}\int_{- \infty}^\infty {\rm d} x\,e^{-isx(k-k')}\right)a^{\dagger}_{s\lambda}(k)\ket{0}\bra{0}a_{s\lambda}(k')
\nonumber \\
&=&\sum_{s, \lambda}\int_{- \infty}^\infty {\rm d} k\int_{- \infty}^\infty {\rm d} k'\sqrt{\frac{|k'|}{|k|}}\delta(k-k')\ket{1_{s\lambda}(k)}\bra{1_{s\lambda}(k')}
\nonumber \\
&=&\sum_{s, \lambda}\int_{- \infty}^\infty {\rm d} k\ket{1_{s\lambda}(k)}\bra{1_{s\lambda}(k)}\, .
\end{eqnarray}
\end{widetext}
Therefore, $\eta\eta^{-1}$ behaves as an identity operator for single excitation states.

\section{Different representations of the free space Hamiltonian}\label{app B}

Using Eq.~\eqref{A dagger bio 1}, the monochromatic photon operators can be represented in terms of the local and bio-local operators as follows
\begin{eqnarray}
a_{s\lambda}^{\dagger}(k) &=& \int_{-\infty}^\infty {\rm d} x \, \sqrt{|k| \over 2 \pi} \, {\rm e}^{ i skx} \, A^{\dagger\,bio}_{s\lambda}(x) \notag \\
&=& \int_{-\infty}^\infty {\rm d} x \, \sqrt{1 \over 2 \pi|k|} \, {\rm e}^{ i skx} \, A^{\dagger}_{s\lambda}(x)\,.
\end{eqnarray}
We can therefore write the free space Hamiltonian in the following representations
\begin{eqnarray}
H_{free}&=&\sum_{s,\lambda}\int_{-\infty}^{\infty}{\rm d}k\, \hbar c k\,a^{\dagger}_{s\lambda}(k)a_{s\lambda}(k)
\nonumber \\
&=&\sum_{s,\lambda}\hbar c\int_{-\infty}^{\infty}{\rm d}x\int_{-\infty}^{\infty}{\rm d}x'\,G(x-x')A_{s\lambda}^{\dagger\, bio}(x)A_{s\lambda}(x')
\nonumber \\
&=&\sum_{s,\lambda}\hbar c\int_{-\infty}^{\infty}{\rm d}x\int_{-\infty}^{\infty}{\rm d}x'\,G(x-x')A_{s\lambda}^{\dagger}(x)A^{bio}_{s\lambda}(x')\,, \notag \\
\end{eqnarray}
where
\begin{equation}
G(y)=\frac{1}{2\pi}\int_{-\infty}^{\infty}{\rm d}k\,ke^{isky}=-is\frac{\rm d}{{\rm d}y}\delta(y)\,
\end{equation}
and is independent of $f(k) = \sqrt{k}$.
Therefore, the free-space Hamiltonian is both bio-Hermitian and Hermitian, i.e. $H^{\dagger\,bio}=H^{\dagger}=H$.

\section{Eq.~(\ref{unitary1}) is not true in general}\label{app C}

To see why Eq.~\eqref{unitary1} is not true in general, we choose an $H_1$ such that at some time $t_1$
\begin{equation}
U^{\dagger}_1(t_1)\,a^{\dagger}_{s\lambda}(x)\,U_1(t_1) = b_1\, a^{\dagger}_{s\lambda}(x,t_1) + b_2\, a_{s\lambda}(x,t_1)
\end{equation}
where $b_1$ and $b_2$ are chosen such that $b_1,b_2 >0$ and $|b_1|^2-|b_2|^2 = 1$.
Taking the one-to-one correspondence $a^{\dagger}_{s\lambda}(x)\rightarrow A^{\dagger}_{s\lambda}(x)$ and $a_{s\lambda}(x)\rightarrow A^{bio}_{s\lambda}(x)$ in the above equation we have
\begin{equation}
U^{\dagger bio}_2(t_1)\,A^{\dagger}_{s\lambda}(x)\,U_2(t_1) = b_1\, A^{\dagger}_{s\lambda}(x,t_1) + b_2\, A^{bio}_{s\lambda}(x,t_1)\,.
\end{equation}
We therefore have
\begin{eqnarray}
\mathcal{R}\left(U^{\dagger}_1(t_1)\,a^{\dagger}_{s\lambda}(x)\,U_1(t_1) \right) &=& b_1 A^{\dagger}_{s\lambda}(x,t_1) + b_2 A_{s\lambda}(x,t_1)
\nonumber \\
&\ne& U^{\dagger bio}_2(t_1)\,A^{\dagger}_{s\lambda}(x)\,U_2(t_1)\,. \notag \\
\end{eqnarray}

\end{document}